%% file: paper.tex
\documentclass[10pt,conference]{IEEEtran}
\IEEEoverridecommandlockouts
\usepackage{cite}
\usepackage{amsmath,amssymb,amsfonts,amsthm}
\usepackage{algorithmic}
\usepackage{graphicx}
\usepackage{textcomp}
\usepackage[usenames,dvipsnames]{xcolor}
\usepackage{hyperref}
\usepackage{mathpartir}
\usepackage{xspace}
\usepackage{listings}
\usepackage{enumitem}
\usepackage{pifont}
\usepackage{caption}
\usepackage{subcaption}
\usepackage[capitalise,nameinlink]{cleveref}
\def\BibTeX{{\rm B\kern-.05em{\sc i\kern-.025em b}\kern-.08em
    T\kern-.1667em\lower.7ex\hbox{E}\kern-.125emX}}

\definecolor{LinkColor}{rgb}{0.55,0.0,0.3}
\definecolor{CiteColor}{rgb}{0.55,0.0,0.3}
\definecolor{HighlightColor}{rgb}{0.0,0.0,0.0}
\definecolor{Gray}{gray}{0.9}
\definecolor{grey}{rgb}{0.5,0.5,0.5}
\definecolor{red}{rgb}{1,0,0}
\definecolor{darkgreen}{rgb}{0.0,0.7,0.0}
\definecolor{backcolor}{rgb}{0.95,0.95,0.95}

\newcommand{\embox}[1]{\mbox{\emph{#1}}}
\newcommand{\C}[1]{\texttt{\small#1}}
\newcommand{\rep}[1]{\overline{#1}}
\newcommand{\eg}{\emph{e.g.}\xspace}
\newcommand{\ie}{\emph{i.e.}\xspace}

\hypersetup{%
  linktocpage=true, pdfstartview=FitV,
  breaklinks=true, pageanchor=true, pdfpagemode=UseOutlines,
  plainpages=false, bookmarksnumbered, bookmarksopen=true, bookmarksopenlevel=3,
  hypertexnames=true, pdfhighlight=/O,
  colorlinks=true,linkcolor=LinkColor,citecolor=CiteColor,
  urlcolor=LinkColor
}

\crefformat{section}{#2\S{}#1#3}
\crefname{figure}{Fig.}{Figures}
\crefname{lstlisting}{Listing}{Listings}
\crefname{table}{Table}{Tables}
\Crefname{line}{Line}{Lines}
\crefname{line}{line}{lines}

\lstdefinestyle{mystyle}{
  backgroundcolor=\color{backcolor},
  keywordstyle=\color{RoyalBlue},
  commentstyle=\color{ForestGreen},
  basicstyle=\ttfamily\footnotesize,
  breakatwhitespace=false,
  breaklines=true,
  captionpos=b,
  keepspaces=true,
  numbers=none,
  numbersep=1pt,
  numberstyle=\scriptsize,
  showspaces=false,
  showstringspaces=false,
  escapechar=$,
}
\lstdefinelanguage{Rust}{
  keywords={as,break,const,continue,crate,else,enum,extern,false,fn,for,if,
  impl,in,let,loop,match,mod,move,mut,pub,ref,return,self,Self,static,struct,
  super,trait,true,type,unsafe,use,where,while,async,await,dyn,abstract,become,
  box,do,final,macro,override,priv,typeof,unsized,virtual,yield,try},
  morecomment=[l]{//},
}

\theoremstyle{remark}
\newtheorem{example}{Example}

\DeclareMathOperator{\argmax}{argmax}

\begin{document}

\title{Concrat: An Automatic C-to-Rust Lock API Translator for Concurrent Programs}

\author{\IEEEauthorblockN{Jaemin Hong}
\IEEEauthorblockA{\textit{School of Computing} \\
\textit{KAIST}\\
Daejeon, South Korea \\
jaemin.hong@kaist.ac.kr}
\and
\IEEEauthorblockN{Sukyoung Ryu}
\IEEEauthorblockA{\textit{School of Computing} \\
\textit{KAIST}\\
Daejeon, South Korea \\
sryu.cs@kaist.ac.kr}
}

\maketitle

\begin{abstract}

  Concurrent programs suffer from data races. To prevent data races, programmers
  use locks. However, programs can eliminate data races only when they acquire and
  release correct locks at correct timing. The lock API of C, in which people
  have developed a large portion of legacy system programs, does not validate
  the correct use of locks. On the other hand, Rust, a recently developed system programming
  language, provides a lock API that guarantees the correct use of locks via
  type checking. This makes rewriting legacy system programs in Rust a promising
  way to retrofit safety into them.  Unfortunately, manual C-to-Rust translation
  is extremely laborious due to the discrepancies between their lock APIs.
  Even the state-of-the-art automatic C-to-Rust translator
  retains the C lock API, expecting developers to replace them with
  the Rust lock API. In this work, we propose an automatic tool to replace the C
  lock API with the Rust lock API. It facilitates C-to-Rust translation of
  concurrent programs with less human effort than the current practice. Our tool
  consists of a Rust code transformer that takes a lock summary as an input and
  a static analyzer that efficiently generates precise lock summaries. We
  show that the transformer is scalable and widely applicable while
  preserving the semantics; it transforms 66 KLOC in 2.6 seconds and successfully
  handles 74\% of real-world programs. We also show that the analyzer is scalable
  and precise; it analyzes 66 KLOC in 4.3 seconds.

\end{abstract}


\input{intro}
\input{background}
\input{overview}
\input{transformation}
\input{analysis}
\input{evaluation}
\input{related}
\input{conclusion}

\section*{Acknowledgment}
This research was supported by National Research Foundation of Korea (NRF)
(Grants 2022R1A2C200366011 and 2021R1A5A1021944), Institute for Information \&
communications Technology Promotion (IITP) grant funded by the Korea government
(MSIT) (2022-0-00460), and Samsung Electronics Co., Ltd (G01210570).

\bibliographystyle{plain}
\bibliography{references}

\end{document}

%% file: intro.tex
\section{Introduction}
\label{sec:intro}

In system programming, concurrency is important yet notoriously difficult to
get right. System software reduces execution time by spawning multiple threads
and splitting tasks. As a drawback, it suffers from various bugs
not existing in the sequential setting: data races, deadlock,
starvation, etc~\cite{concurrency-bugs}.

Data races are the most common category of concurrency
bugs~\cite{concurrency-bugs}. It happens when multiple threads read and write the same
memory address simultaneously. Data races lead system programs to exhibit not
only unpleasant malfunctions but also critical security
vulnerabilities~\cite{10.1145/2103799.2103805}.

Among synchronization mechanisms to avoid data races, locks are the most
widely-used one. Each thread acquires and releases a lock before and after
accessing shared data. This simplicity has facilitated the adoption of locks in
diverse system software. Unfortunately, locks prevent data races only when they
are used correctly. Programmers may acquire wrong locks, acquire locks too
late, or release locks too early, thereby failing to eliminate data races.

C, in which people have developed a significant portion of system programs,
burdens programmers with the validation of correct lock use. The most popular
lock API of C, pthreads~\cite{pthreads}, does not automatically check whether
programs use locks correctly. Developers often fail to recognize incorrectly
used locks in their programs, and C programs thus have suffered from data races.

Rust~\cite{matsakis2014rust,rust}, a recently developed system programming language, provides a lock API
guaranteeing \emph{thread safety}, \ie, the absence of data races, in \C{std::sync}
of its standard library~\cite{std-sync}. The combination of the ownership type
system of Rust and the carefully designed API allows the type checker to
validate the correct use of locks at compile time~\cite{rustbelt}. The API is
different from the C lock API not only syntactically, \eg, in names of
functions, but also semantically. For instance, the Rust lock API requires
programs to explicitly describe which lock protects which data, while the C lock
API does not.

Thanks to the thread-safe lock API of Rust, rewriting legacy concurrent C programs in
Rust is a promising approach to secure safety. It can reveal unknown data races
and allow developers to make fixes. In addition, rewriting in Rust
prevents introducing new data races while adding new features.

Noticing this potential, programmers have begun to rewrite concurrent
programs in Rust. Mozilla developed Servo, a web browser written in Rust, and
has replaced modules of Firefox with those of Servo. Its developers said that
Rust’s thread safety significantly helped implement concurrent
renderers correctly~\cite{firefox}. People also adopt Rust into operating
systems, in which concurrency is extremely important. The next release of the
Linux kernel will support Rust code~\cite{linux}. Android and Fuchsia
implementations also use Rust~\cite{android,fuchsia}.

Reimplementing concurrent programs in Rust is, however, labor-intensive if done
manually. The discrepancies between the lock APIs of C and Rust hinder
programmers from mechanical rewriting. They have to understand the use of the C
lock API in original programs and restructure the programs to express the intended
logic with the Rust lock API. It poses the necessity for a tool for automatic
C-to-Rust translation.

The state-of-the-art C-to-Rust translator, C2Rust~\cite{c2rust}, is
still far from alleviating the burden on programmers. The translation of C2Rust
is completely syntactic and generates Rust code using the C lock API. Programmers
are expected to replace the C lock API in C2Rust-generated code with the Rust lock
API for thread safety. While being better than nothing, the benefit from
C2Rust’s syntactic translation is marginal.

In this work, we propose an automatic tool to replace the C lock API with
the Rust lock API. Specifically, we make the following contributions:
\begin{itemize}
\item We identify the problem of replacing the C lock API with
the Rust lock API as the first step toward automatic C-to-Rust
translation of concurrent programs~(\cref{sec:overview}).
\item We propose an automatic Rust code transformer that takes
C2Rust-generated code and its lock summary as inputs
and replaces the C lock API with the Rust lock API~(\cref{sec:transformation}).
\item We propose a static analyzer that efficiently generates a precise
lock summary by combining bottom-up dataflow analysis and top-down
data fact propagation~(\cref{sec:analysis}).
\item We build Concrat ({\bf Con}current-{\bf C} to {\bf R}ust {\bf A}utomatic {\bf T}ranslator)
by combining C2Rust, the static analyzer, and the transformer.
Our evaluation shows that the transformer efficiently and correctly transforms
real-world programs and the analyzer outperforms the
state-of-the-art static analyzer in terms of both speed and
precision. Specifically, they transform and analyze 66 KLOC in 2.6 seconds and
4.3 seconds, respectively, and translate 74\% of real-world programs to
compilable code~(\cref{sec:evaluation}).
\end{itemize}
We discuss related work~(\cref{sec:related}) and conclude~(\cref{sec:conclusion}).
Our implementation and evaluation data are publicly
available~\cite{hong_jaemin_2023_7573490}.

%% file: background.tex
\section{Background: Lock APIs of C and Rust}
\label{sec:background}

\subsection{Lock API of C}
\label{sec:background-c}

The most widely used C lock API, pthreads~\cite{pthreads}, provides three types
of lock: mutexes, read-write locks, and spin locks. While all of them are within the
scope of this work, we mainly discuss mutexes throughout the paper. The others
are similar to mutexes and briefly discussed in \cref{sec:implementation}.

The API provides the \C{pthread\_mutex\_lock} and \C{pthread\_mutex\_unlock} functions,
each of which takes a pointer to a lock; the former acquires the
lock, and the latter releases the lock. Locks
are used together with shared
data. C programs have two common patterns to organize locks and shared data:
\emph{global} and \emph{struct}~\cite{goblint}.

\subsubsection{Global}
\label{sec:global}

Both data and lock are global variables.
\begin{lstlisting}[language=C]
int n = ...; pthread_mutex_t m = ...;
void inc() {
    pthread_mutex_lock(&m); n += 1;
    pthread_mutex_unlock(&m);
}
\end{lstlisting}
The global variable \C{n} is a shared integer and \C{m} is a lock. Each thread
must hold \C{m} when accessing \C{n}, \ie, \C{m} protects \C{n}. Thus,
\C{inc} acquires and releases \C{m} before and after increasing \C{n}.



\subsubsection{Struct}
\label{sec:struct}

Both data and lock are fields of the same struct.
\begin{lstlisting}[language=C]
struct s { int n; pthread_mutex_t m; };
void inc(struct s *x) {
    pthread_mutex_lock(&x->m); x->n += 1;
    pthread_mutex_unlock(&x->m);
}
\end{lstlisting}
The lock stored in the field \C{m} of a struct \C{s} value protects the integer
stored in the field \C{n} of the same struct value. Each thread must hold
\C{x->m} when accessing \C{x->n}.


\subsection{Data Races in C}
\label{sec:background-bug}

The C lock API does not guarantee whether programs use locks correctly.  Data
races may occur by mistake despite the use of locks.  There are two major
reasons for data races: \emph{data-lock mismatches} and \emph{flow-lock
mismatches}.

\subsubsection{Data-Lock Mismatch}

A data-lock mismatch is an acquisition of an incorrect lock when accessing
shared data. See the following where \C{m1} protects \C{n1} and
\C{m2} protects \C{n2}:
\begin{lstlisting}[language=C]
pthread_mutex_lock(&m2); n1 += 1;
pthread_mutex_unlock(&m2);
\end{lstlisting}
All the other parts of the program acquire \C{m1} when accessing \C{n1}.
However, the above code has a bug: it acquires \C{m2}, instead of \C{m1}, when
accessing \C{n1}. This allows multiple threads to access \C{n1} simultaneously,
thereby incurring a data race.

\subsubsection{Flow-Lock Mismatch}

A flow-lock mismatch is an acquisition of a lock at an incorrect program point.
Consider the following program, where \C{m} protects \C{n}:
\begin{lstlisting}[language=C]
void f1() { n += 1; pthread_mutex_lock(&m); ... }
void f2() { ... pthread_mutex_unlock(&m); n += 1; }
\end{lstlisting}
The function \C{f1} accesses \C{n} before acquiring \C{m}, and \C{f2} accesses
\C{n} after releasing \C{m}. Both functions are buggy as they allow accesses to
the shared data when the lock is not held.

\subsection{Lock API of Rust}
\label{sec:background-rust}

The Rust lock API guarantees the correct use of locks~\cite{rustbelt}. The API makes
two kinds of relation explicit: \emph{data-lock relations}, \ie, which
lock protects which data; \emph{flow-lock relations}, \ie, which lock is held at
which program point. It naturally prevents both data-lock mismatches and
flow-lock mismatches.

Rust makes the data-lock relation explicit by coupling a lock with shared data.
A Rust lock is a C lock plus shared data; it can be considered a protected
container for shared data. The type of a lock is \C{Mutex<T>}, where \C{T} is
the type of the protected data~\cite{mutex}. A program can create a lock as
follows:
\begin{lstlisting}[language=Rust]
static m: Mutex<i32> = Mutex::new(0);
\end{lstlisting}
making \C{m} a lock initially containing \C{0}.
The coupling of data and a lock prevents data-lock mismatches. When
accessing shared data, threads acquire the lock coupled with the data.

Rust makes the flow-lock relations explicit by introducing the notion of a
\emph{guard}.  Threads need a guard to access the in-lock data. A guard is a
special kind of pointer to the in-lock data. The only way to create a guard is to
acquire a lock. The \C{lock} method of a lock produces a guard as a return
value.  Threads can access the protected data by dereferencing the guard.  When
a thread wants to release a lock, it drops, \ie, deallocates, the guard by
calling \C{drop}. A predefined drop handler attached to the guard automatically
releases the lock.
The following shows the process from construction to destruction of a guard:
\begin{lstlisting}[language=Rust]
let mut g = m.lock().unwrap(); *g += 1; drop(g);
\end{lstlisting}
Because \C{lock} returns a wrapped guard, \C{unwrap()} is required.
The \C{unwrap} call fails and makes the current thread panic when a thread
previously holding the lock has panicked before releasing it.  Otherwise,
\C{unwrap} returns the guard.

With the help of Rust's ownership type system~\cite{rustbelt}, guards prevent flow-lock mismatches.
In Rust, each function can use only the variables it owns. A function
owns a variable after initializing it and loses the ownership after passing
the variable to another function as an argument. The type checker detects every
flow-lock mismatch at compile time by tracking the ownership of guards.
Consider the following buggy code:
\begin{lstlisting}[language=Rust]
fn f1() {let mut g; *g += 1; g = m.lock().unwrap();}
fn f2() { let mut g; ... drop(g); *g += 1; }
\end{lstlisting}
Because \C{f1} uses \C{g} before owning it and \C{f2} uses \C{g} after
losing the ownership, the type checker rejects both functions.

%% file: overview.tex
\section{Motivation}
\label{sec:overview}

\begin{figure}[t]
  \centering
  \includegraphics[width=0.45\textwidth]{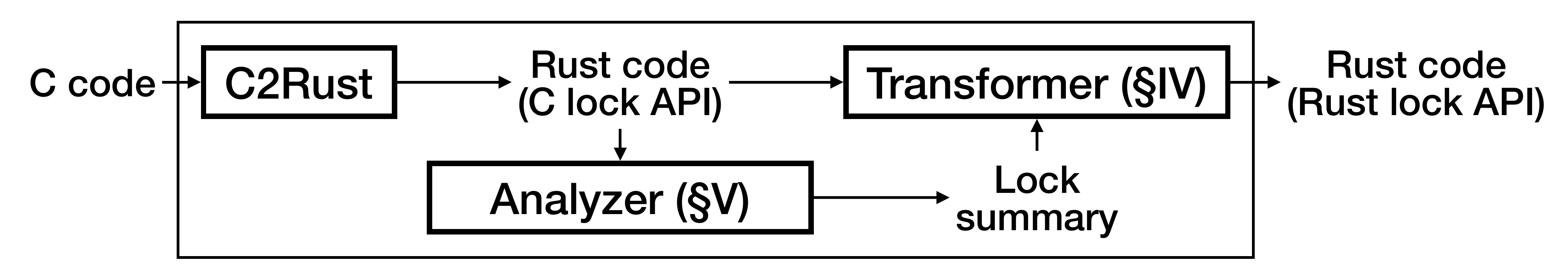}
\vspace*{-0.5em}
  \caption{Workflow of Concrat}
  \label{fig:concrat}
\vspace*{-2em}
\end{figure}

C-to-Rust translation should ideally replace the C lock API with the Rust lock API
because implementing concurrent programs in Rust benefits mostly from the thread
safety guaranteed by the Rust lock API.  From this perspective,
C2Rust~\cite{c2rust}, the state-of-the-art C-to-Rust automatic translator, is
unsatisfactory.  It syntactically translates C code to Rust code. As a result, the C
lock API remains in C2Rust-generated code. For example, the result of
translating the first C code snippet in \cref{sec:background-c} with C2Rust is
as follows:
\begin{lstlisting}[language=Rust]
static mut n: i32 = ...;
static mut m: pthread_mutex_t = ...;
fn inc() {
    pthread_mutex_lock(&mut m); n += 1;
    pthread_mutex_unlock(&mut m);
}
\end{lstlisting}

To narrow the gap between C2Rust-generated code and desirable Rust
code, we present the problem of \emph{replacing the C lock API in C2Rust-generated
code with the Rust lock API}. Solving this problem will significantly reduce
programmers' burden when rewriting legacy concurrent programs in Rust.

Our solution is Concrat, an automatic C-to-Rust translator for concurrent
programs. \cref{fig:concrat} shows the workflow of Concrat. It takes C code and
translates it using C2Rust. A static analyzer generates a lock summary of the
Rust code (\cref{sec:analysis}). A transformer converts the C lock API to the
Rust lock API using the summary (\cref{sec:transformation}). Concrat outputs the
transformed code.

%% file: transformation.tex
\section{Transformation of C2Rust-Generated Code}
\label{sec:transformation}

\begin{figure}[t]
\begin{lstlisting}[language=Rust,numbers=left,numbersep=1pt,numberstyle=\tiny,
caption=Before transformation,label=lst:before]
static mut n: i32 = 0;
static mut m: pthread_mutex_t = ...;
struct s { n: i32, m: pthread_mutex_t }
fn f() {
    pthread_mutex_lock(&mut m);
    n += 1; pthread_mutex_unlock(&mut m); }
fn unlock() { pthread_mutex_unlock(&mut m); }
fn lock() {
    pthread_mutex_lock(&mut m); }
fn g() {
    lock();
    n += 1; unlock(); }
fn foo() {
    n += 1; // safe for some reason
    pthread_mutex_lock(&mut m);
    n += 1;$\label{line:held1}$
    pthread_mutex_unlock(&mut m); }$\label{line:held2}$
\end{lstlisting}
\vspace*{-1em}
\begin{lstlisting}[caption=Lock summary,label=lst:summary]
{ "global_lock_map": { "n": "m" },
  "struct_lock_map": { "s": { "n": "m" } },
  "function_map": {
    "unlock": {
      "entry_lock": ["m"], "return_lock": [], ... },
    "lock": {
      "entry_lock": [], "return_lock": ["m"], ... },
    "foo": { ... "lock_line": { "m": [16, 17] }, },
    ... }}
\end{lstlisting}
\vspace*{-1em}
\begin{lstlisting}[language=Rust,numbers=left,numbersep=1pt,numberstyle=\tiny,
caption=After transformation,label=lst:after]
struct mData { n: i32 }$\label{line:global1}$
static mut m: Mutex<mData> = Mutex::new(mData{n:0});$\label{line:global2}$
struct smData {n:i32}  struct s {m:Mutex<smData>}$\label{line:struct}$
fn f() { let mut m_guard;$\label{line:access1}$
    m_guard = m.lock().unwrap();
    (*m_guard).n += 1; drop(m_guard); }$\label{line:access2}$
fn unlock(m_guard: MutexGuard<i32>) {drop(m_guard);}$\label{line:unlock}$
fn lock() -> MutexGuard<i32> { let mut m_guard;$\label{line:lock1}$
    m_guard = m.lock().unwrap(); m_guard }$\label{line:lock2}$
fn g() { let mut m_guard;$\label{line:call1}$
    m_guard = lock();
    (*m_guard).n += 1; unlock(m_guard); }$\label{line:call2}$
fn foo() { let mut m_guard;$\label{line:getmut1}$
    m.get_mut().n += 1; // safe for some reason$\label{line:noguard}$
    m_guard = m.lock().unwrap();
    (*m_guard).n += 1;$\label{line:guard1}$
    drop(m_guard); }$\label{line:getmut2}$$\label{line:guard2}$
\end{lstlisting}
\vspace*{-3em}
\end{figure}

This section proposes an automatic Rust code transformer that takes
C2Rust-generated code and its lock summary as inputs and replaces
the C lock API with the Rust lock API.
We first describe the contents
of a lock summary (\cref{sec:lock-summary}) and then show how the transformer
replaces the C lock API with the Rust lock API using the summary
(\cref{sec:transformer}).

\subsection{Lock Summary}
\label{sec:lock-summary}

\newcommand{\glm}{\C{global\_lock\_map}\xspace}
\newcommand{\slm}{\C{struct\_lock\_map}\xspace}
\newcommand{\funcm}{\C{function\_map}\xspace}
\newcommand{\el}{\C{entry\_lock}\xspace}
\newcommand{\rl}{\C{return\_lock}\xspace}
\newcommand{\lline}{\C{lock\_line}\xspace}

A program's lock summary abstracts its data-lock and flow-lock relations.
A lock summary is a JSON file containing three maps: \glm,
\slm, and \funcm. The first two represent data-lock relations, and the last
one represents flow-lock relations. \cref{lst:before} is C2Rust-generated
code, and \cref{lst:summary} is its lock summary. We describe each component of
the summary in detail.

\subsubsection{Global Lock Map}

\glm expresses data-lock relations of the global pattern (\cref{sec:global}).
It is a map from a global variable to the lock variable protecting it.  The
summary states that \C{n} is protected by \C{m}.

\subsubsection{Struct Lock Map}

\slm keeps data-lock relations of the struct pattern (\cref{sec:struct}).
It maps a struct type name to its summary, which maps a field
to the lock field protecting it. The summary states that
the field \C{n} of the struct type \C{s} is protected by the field \C{m}.

\subsubsection{Function Map}

\funcm expresses the flow-lock relation. It maps a function name to its
summary, which consists of \el, \rl, and \lline.
Locks are represented by symbolic paths. For example, \C{m}
and \C{x.m} are locks, where the type of the variable \C{x} is \C{s}.

\paragraph{Entry Lock}

\el is a list of locks that are always held at the entry of the function. The
summary states that \C{m} is always held at the entry of \C{unlock}.

\paragraph{Return Lock}

\rl is a list of locks that are always held at the return of the function. The
summary states that \C{m} is always held at the return of \C{lock}.

\paragraph{Lock Line}

\lline is a map from a lock to a list of lines in the function where the lock
is held. The summary states that \C{m} is held in \cref{line:held1,line:held2}.

\subsection{Transformation}
\label{sec:transformer}

The transformer produces \cref{lst:after} by replacing the C lock API in
\cref{lst:before} with the Rust lock API. Note that each line of
\cref{lst:after} corresponds to the same line of \cref{lst:before}.
We explain the transformation line-by-line.

\begin{itemize}[leftmargin=*]
  \item \Crefrange{line:global1}{line:global2}:
    We check \glm to identify locks and variables they protect.
    We define a new struct containing variables protected by a certain lock and
    replace the original C lock with a Rust lock containing a struct value.
  \item \Cref{line:struct}: Similar to the above, but using \slm, instead of
    \glm.
  \item \Crefrange{line:access1}{line:access2}: We define an uninitialized guard
    variable at the beginning of each function using the guard. Each
    \C{pthread\_mutex\_lock} and \C{pthread\_mutex\_unlock} call is
    syntactically transformed into a \C{lock} method call and a \C{drop}
    function call, respectively. Note that the name of a guard is syntactically determined from
    the name of the lock according to a predefined rule.
    We replace each expression accessing protected data with an
    expression dereferencing a guard, whose lock name is found in \glm or \slm,
    depending on the access path.
  \item \Cref{line:unlock}:
    We make a function take a guard as an argument if its \el is nonempty.
  \item \Crefrange{line:lock1}{line:lock2}:
    We make a function return a guard if its \rl is nonempty. If there are
    multiple return guards or the original return value, tuples are
    constructed.
  \item \Crefrange{line:call1}{line:call2}:
    We add a guard as an argument to a call to a function with nonempty \el. We assign
    the return value of a function with nonempty \rl to a guard variable.
  \item \Crefrange{line:getmut1}{line:getmut2}:
    Even when \C{m} protects \C{n}, some accesses to \C{n} may not hold \C{m}
    because the developer thinks that \C{n} is never concurrently accessed by
    other threads in those specific lines. For this reason, we cannot blindly
    replace all the accesses to protected data with guard dereference. We need
    to figure out whether a certain guard exists in each line by checking
    \lline. Since the summary states that \C{m} is held only in
    \cref{line:guard1,line:guard2}, the access in \cref{line:noguard} uses the
    \C{get\_mut} method, instead of the guard. The method returns a pointer to
    the in-lock data.

    Note that the use of \C{get\_mut} relies on that \C{m} is defined mutable.
    If \C{m} is immutable, the type checker disallows calling \C{get\_mut}.  In
    Rust, mutable global variables are discouraged~\cite{static-mut}.
    Reference-counted types, \C{Rc}~\cite{rc} (in the sequential setting) and
    \C{Arc}~\cite{arc} (in the concurrent setting), should replace mutable
    global variables. This makes \C{get\_mut} succeed if the reference count
    equals one and panic otherwise, consequently preventing data races at run
    time even when the developer's assumption is wrong. Automatically replacing
    mutable global variables with \C{Rc} and \C{Arc} is beyond the scope of this
    work.
\end{itemize}

The transformed code looks similar to human-written code because the
transformer utilizes code patterns that real-world Rust programmers use, e.g.,
putting fields into structs protected by locks, storing guards in variables,
passing guards as arguments, and returning guards from functions. Still, there
are some discrepancies: humans may prefer wrapping guards in structs and
defining their methods instead of functions taking guards; they often omit
\C{drop} calls at the end of a function, which can be automatically inserted by the
compiler.

%% file: analysis.tex
\newcommand{\dvar}[3]{\mathsf{\small #1}^\textsf{\tiny #2}_{#3}}
\newcommand{\entryn}{\ensuremath{\mathsf{\small entry}}\xspace}
\newcommand{\retn}{\ensuremath{\mathsf{\small ret}}\xspace}
\newcommand{\alias}{\ensuremath{\embox{alias}}\xspace}

\section{Summary Generation via Static Analysis}
\label{sec:analysis}

In this section, we propose a static analysis to automatically generate lock
summaries required by the transformer.
%
The analysis must precisely determine the flow-lock relation to lead the
transformer to produce compilable code. If a summary contains an imprecise
flow-lock relation, the transformed code may be uncompilable due to the use of
unowned guards.
On the other hand,
an imprecise
data-lock relation does not hinder the transformed code from being compiled.  If
the analysis fails to find that \C{m} protects \C{n}, \C{n} will not be a field
of a struct protected by \C{m}.  If the analysis
incorrectly concludes that \C{m} protects \C{n}, \C{n} will be accessed via
\C{get\_mut}. Both kinds of code are unideal but compilable
and preserve the original semantics.
We thus focus on designing an analysis that precisely computes the flow-lock
relation.

The key intuition behind our analysis design is that the precision of the
analysis does not need to exceed that of the type checker.
Consider the following example:
\begin{lstlisting}[language=Rust]
if b { pthread_mutex_lock(&m); } ...
if b { pthread_mutex_unlock(&m); }
\end{lstlisting}
Even when the analysis is precise enough to track the path-sensitive use of
locks, the transformed code is uncompilable:
\begin{lstlisting}[language=Rust]
let mut m_guard;
if b { m_guard = m.lock().unwrap(); } ...
if b { drop(m_guard); }
\end{lstlisting}
Since type checking is path-insensitive, it considers \C{m\_guard} possibly
uninitialized in the last line. This shows that a path-insensitive analysis is
enough. Similarly, our analysis can be context-insensitive as the type
checker is context-insensitive.

This intuition makes our analysis distinct from existing techniques: it is
tailored to efficiently generate precise summaries for the code transformation
by aiming the same precision as the type checker. Existing ones are either too
imprecise or too precise. Some overapproximate the behavior of a program too
much, so using their results as summaries would make the transformed code
uncompilable.  Some unnecessarily adopt rich techniques
to make their results precise, thereby failing to finish the
analyses in a reasonable amount of time.

Note that aiming the same precision as the type checker does not mean that we
repeat the work of the type checker. While the goal of the type checker is to
validate the use of guards, our goal is to infer the use of guards, which is
more difficult. Specifically, the type checker takes code that already has
guards and checks whether it uses guards properly in terms of ownership, but
our analyzer takes code without any guards and reconstructs the flow
of guards to determine whether each function needs to take or return certain
guards.

Since guards are more concrete than information that certain locks
are held, guards often make our explanation intuitive.
Thus, we sometimes use guards in the explanation although
the code being analyzed does not have any guards.
The existence of a guard at a certain program point is equivalent to
the corresponding lock always being held at the program point, and the term
guard is exchangeable for the term \emph{held lock}.

Our analysis consists of four phases: call graph construction
(\cref{sec:call-graph}), bottom-up dataflow analysis (\cref{sec:bottom-up}),
top-down data fact propagation (\cref{sec:top-down}), and data-lock relation
identification (\cref{sec:data-lock}). The call graph is required for both
bottom-up analysis and top-down propagation. The bottom-up analysis and the
top-down propagation collectively compute the flow-lock relation. Using the
flow-lock relation, the last phase computes the data-lock relation.

\vspace{-0.25em}
\subsection{Call Graph Construction}
\label{sec:call-graph}

We draw call graphs by collecting the function names called
in each function, without expensive control flow analysis. The
drawback is that the call graph misses edges created by function pointers.
However, the number of such edges is usually small because function pointers
are rarely used in practice, and the subsequent analyses remain precise enough.
Each node is a user-defined function; all the
library functions are excluded from the graph. Therefore, each leaf node calls
zero or more library functions but no user-defined functions.

We identify all the strongly connected components in the call graph to find
mutually recursive functions, which need special treatment during the bottom-up
analysis. We create a \emph{merged} version of the call graph by merging each
strongly connected component into a single node. We keep both original and
merged call graphs to use the former for the top-down propagation and the latter
for the bottom-up analysis.

\vspace{-0.25em}
\subsection{Bottom-Up Dataflow Analysis}
\label{sec:bottom-up}

The goal of the bottom-up analysis is to identify the \emph{minimum entry lock
set} (MELS) and the \emph{minimum return lock set} (MRLS) of each function.
They are locks that must be held at the entry and the return, respectively.
To compute the MELS and MRLS of each function, we perform two
dataflow analyses on each function: \emph{live guard analysis} (LGA) and
\emph{available guard analysis} (AGA). LGA computes MELSs, and AGA computes
MRLSs. We need the control flow graph of each function for the
analyses. The nodes are statements of the
function, with two special nodes, \entryn and \retn, which
denote the entry and the return, respectively.

We traverse the merged call graph in post order to find the analysis target. It
allows us to analyze leaf nodes first (\cref{sec:leaf}) and then use their
results to analyze internal nodes (\cref{sec:internal}). Each node contains a
single function or a set of mutually recursive functions. We discuss the
analysis of non-recursive functions first and recursive functions afterward
(\cref{sec:recursive}).

\subsubsection{Leaf Node}
\label{sec:leaf}

\paragraph{Live Guard Analysis}

The goal of LGA is to compute MELSs. It is similar to the well-known live
variable analysis~\cite{data-flow-analysis}. Just like that the live variable
analysis computes variables to be used in the future, LGA computes guards to be
consumed by \C{pthread\_mutex\_unlock} in the future. Live guards at the entry
of a function are the MELS of the function.

The analysis is a backward may analysis. Each \C{pthread\_mutex\_unlock} call,
which consumes a guard, generates a guard.  Each \C{pthread\_mutex\_lock} call,
which produces a guard, kills a guard.  The dataflow equations are defined as
follows:
\vspace{-0.25em}
\[
\small
\begin{array}{r@{~}c@{~}l}
  \dvar{In}{L}{s} &=&
    (\dvar{Out}{L}{s}-\dvar{Kill}{L}{s})\cup\dvar{Gen}{L}{s} \\
  \dvar{Out}{L}{s} &=&
    \begin{cases}
      \emptyset & \text{if}\ s=\retn \\
      \bigcup_{t\in\dvar{\scriptsize Succ}{}{s}}\dvar{In}{L}{t} & \text{otherwise}
    \end{cases} \\[8pt]
  \dvar{Gen}{L}{s} &=&
    \begin{cases}
      \{p\} & \text{if}\ s=\C{pthread\_mutex\_unlock}(p) \\
      \emptyset & \text{otherwise}
    \end{cases} \\[8pt]
  \dvar{Kill}{L}{s} &=&
    \begin{cases}
      \{p\} & \text{if}\ s=\C{pthread\_mutex\_lock}(p) \\
      \emptyset & \text{otherwise}
    \end{cases} \\
  \dvar{MELS}{}{} &=& \dvar{In}{L}{\entryn}
\end{array}
\vspace{-0.25em}
\]
where $s$ and $t$ range over statements; $p$ ranges over paths; $\dvar{Succ}{}{s}$ denotes
the set of every successor of $s$.

\begin{example} The MELS of the following function is $\{\C{m}\}$:
\begin{lstlisting}[language=Rust]
fn unlock() { pthread_mutex_unlock(&mut m); }
\end{lstlisting}
\vspace{-0.5em}
\end{example}

\begin{example} The MELS of the following function is $\{\C{m}\}$:
\begin{lstlisting}[language=Rust]
fn may_unlock() {
    if ... { pthread_mutex_unlock(&mut m); }
}
\end{lstlisting}
\vspace{-0.5em}
We get $\{\C{m}\}$ by $\{\C{m}\} \cup \emptyset$ because LGA is
a may analysis. If it was a must analysis, MELS would be $\emptyset$, making the
function take no guard after the transformation. Then, the function is
uncompilable as it drops an unexisting guard.
\end{example}

\paragraph{Available Guard Analysis}

The goal of AGA is to compute MRLSs. It is similar to the well-known
available expression analysis~\cite{data-flow-analysis}. Just like that the
available expression analysis identifies expressions whose values have been
computed in the past, AGA identifies guards constructed by
\C{pthread\_mutex\_lock} in the past.  Available guards at the return of a
function are the MRLS of the function.

The analysis is a forward must analysis. Each \C{pthread\_mutex\_lock} call
generates a guard, and each \C{pthread\_mutex\_unlock} call kills a guard. The
dataflow equations are defined as follows:
\vspace{-0.5em}
\[
\small
\begin{array}{r@{~}c@{~}l}
  \dvar{Out}{A}{s} &=&
    (\dvar{In}{A}{s}-\dvar{Kill}{A}{s})\cup\dvar{Gen}{A}{s} \\
  \dvar{In}{A}{s} &=&
    \begin{cases}
      \dvar{MELS}{}{} & \text{if}\ s=\entryn \\
      \bigcap_{t\in\dvar{\scriptsize Pred}{}{s}}\dvar{Out}{A}{t} & \text{otherwise}
    \end{cases} \\[8pt]
  \dvar{Gen}{A}{s} &=&
    \begin{cases}
      \{p\} & \text{if}\ s=\C{pthread\_mutex\_lock}(p) \\
      \emptyset & \text{otherwise}
    \end{cases} \\[8pt]
  \dvar{Kill}{A}{s} &=&
    \begin{cases}
      \{p\} & \text{if}\ s=\C{pthread\_mutex\_unlock}(p) \\
    \emptyset & \text{otherwise}
  \end{cases} \\
  \dvar{MRLS}{}{} &=& \dvar{Out}{A}{\retn}
\end{array}
\vspace{-0.5em}
\]
where $\dvar{Pred}{}{s}$ denotes the set of every predecessor of $s$.

\begin{example}
The MRLS of the following function is $\{\C{m}\}$:
\begin{lstlisting}[language=Rust]
fn lock() { pthread_mutex_lock(&mut m); }
\end{lstlisting}
\vspace{-0.5em}
\end{example}

\begin{example}
The MRLS of the following function is $\emptyset$:
\begin{lstlisting}[language=Rust]
fn may_lock() {
    if ... { pthread_mutex_lock(&mut m); }
}
\end{lstlisting}
\vspace{-0.5em}
We get $\emptyset$ by $\{\C{m}\} \cap \emptyset$ because AGA is
a must analysis. If it was a may analysis, MRLS would be $\{\C{m}\}$, making the
function return a guard after the transformation, which is
uncompilable as it returns a possibly uninitialized guard.
\end{example}

\begin{example}
Both MELS and MRLS of the following are $\{\C{m}\}$:
\begin{lstlisting}[language=Rust]
fn unlock_and_lock() {
    if ... {
        pthread_mutex_unlock(&mut m); ...
        pthread_mutex_lock(&mut m);
    }
}
\end{lstlisting}
\vspace{-0.5em}
We get $\{\C{m}\}$ as the MRLS by intersecting $\{\C{m}\}$ and $\{\C{m}\}$. It
is because setting $\dvar{In}{A}{\entryn}$ to $\dvar{MELS}{}{}$ allows
\C{m} to be available even in the path where the condition is false. If
$\dvar{In}{A}{\entryn}$ was $\emptyset$, the MRLS would be $\emptyset$, making
the transformed function not return an existing guard.
\end{example}

\subsubsection{Internal Node}
\label{sec:internal}

Analysis of internal nodes should consider the MELSs and MRLSs of
callees.  A function with a nonempty MELS consumes guards and acts like
\C{pthread\_mutex\_unlock}. A function with a nonempty MRLS produces guards,
like \C{pthread\_mutex\_lock}. Thus, during LGA, calling
$f$ kills $\dvar{MRLS}{}{f}$ and generates $\dvar{MELS}{}{f}$, and during AGA,
calling $f$ kills $\dvar{MELS}{}{f}$ and generates $\dvar{MRLS}{}{f}$.

\begin{example}
The MELS of \C{unlock2} is $\{\C{m}\}$.
\begin{lstlisting}[language=Rust]
fn unlock() { pthread_mutex_unlock(&mut m); }
fn unlock2() { unlock(); }
\end{lstlisting}
\vspace{-0.5em}
\end{example}

When structs are involved, we need to consider aliasing through argument
passing. A caller and a callee represent the same lock with different paths if
the path being an argument is different from the name of the corresponding
parameter. Unless we recompute paths to reflect aliasing, the analyses produce
incorrect results.

In this regard, we define \alias, which recomputes paths:
\vspace{-0.25em}
\[
\small
\alias(p,[x_1,\cdots\!,x_n],[e_1,\cdots\!,e_n])=
\begin{cases}
  e_i.p' & \text{if}\ p=x_i.p' \\
  p & \text{otherwise} \\
\end{cases}
\vspace{-0.25em}
\]
It takes a path, a parameter list, and an argument list. If the path has one of
the parameters as a prefix, \alias replaces the prefix with the corresponding argument.
Otherwise, the path remains the same. For example,
$\alias(\C{a.m},[\C{a}],[\C{b}])$ equals \C{b.m}.
Since we have a set of paths, we extend the definition of \alias to recompute
each path in a given set:
\[
\small
\alias(\{\cdots\!,p,\cdots\!\},\rep{x},\rep{e})=\{\cdots\!,\alias(p,\rep{x},\rep{e}),\cdots\!\}
\vspace{-0.25em}
\]
An overlined symbol denotes a list.
We revise our dataflow equations to handle user-defined function calls
correctly:
\[
\small
\begin{array}{l@{~}c@{~}c@{~}c@{~}c@{~}l}
  \text{If}\ s=f(\rep{e}), &
  \dvar{Gen}{L}{s} &=& \dvar{Kill}{A}{s} &=& \alias(\dvar{MELS}{}{f},\dvar{Params}{}{f},\rep{e}) \\
  & \dvar{Kill}{L}{s} &=& \dvar{Gen}{A}{s} &=& \alias(\dvar{MRLS}{}{f},\dvar{Params}{}{f},\rep{e}) \\
\end{array}
\vspace{-0.25em}
\]
where $\dvar{Params}{}{f}$ denotes the parameter list of $f$.

\begin{example}
The MELS of \C{lock\_and\_unlock} is $\emptyset$.
\vspace{-0.25em}
\begin{lstlisting}[language=Rust]
fn unlock(a: *mut s) {
    pthread_mutex_unlock(&mut (*a).m);
}
fn lock_and_unlock(b: *mut s) {
    pthread_mutex_lock(&mut (*b).m); unlock(b);
}
\end{lstlisting}
\vspace{-0.5em}
While the MELS of \C{unlock} is $\{\C{a.m}\}$, the \C{unlock} call in
\C{lock\_and\_unlock} generates \C{b.m}, which is
killed by the preceding \C{pthread\_mutex\_lock} call.
\end{example}

\subsubsection{Recursive Function}
\label{sec:recursive}

Analysis of a recursive function is challenging because it requires the MELS and
MRLS of the function being analyzed. Our solution is an iterative
analysis.

In the beginning, we have no information and set the MELS and MRLS to the
bottom values: $\dvar{MELS}{}{}=\emptyset$ and $\dvar{MRLS}{}{}=\mathcal{L}$, the set of every possible lock path.
For the MELS, $\emptyset$ is the bottom because LGA is a may analysis. On the
other hand, $\mathcal{L}$ is the bottom
for the MRLS because AGA is a must analysis.

We iteratively find a fixed point to compute the correct MELS and MRLS.  We
analyze the function with the MELS and MRLS we have. After the analysis, we
update them with the result of the analysis. We repeat this until no change.

\begin{example}
The MELS of the following function is $\{\C{m}\}$.
\vspace{-0.25em}
\begin{lstlisting}[language=Rust]
fn unlock(n: i32) {
    if n <= 0 { pthread_mutex_unlock(&mut m); }
    else { unlock(n - 1); }
}
\end{lstlisting}
\vspace{-0.5em}
The first iteration gives us $\dvar{MELS}{}{}=\{\C{m}\}$ by
$\{\C{m}\} \cup\emptyset$. The second iteration produces the same result by
$\{\C{m}\} \cup\{\C{m}\}$ and reaches a fixed point.
\end{example}

\begin{example}
The MRLS of the following function is $\{\C{m}\}$.
\vspace{-0.25em}
\begin{lstlisting}[language=Rust]
fn lock(n: i32) {
    if n <= 0 { pthread_mutex_lock(&mut m); }
    else { lock(n - 1); }
}
\end{lstlisting}
\vspace{-0.5em}
The first iteration makes $\dvar{MRLS}{}{}=\{\C{m}\}$ by
$\{\C{m}\} \cap\mathcal{L}$. The second iteration computes
$\{\C{m}\} \cap \{\C{m}\}$, reaching a fixed point.
\end{example}

The iteration is guaranteed to terminate if $\mathcal{L}$ is finite.
During the iteration, MELS can only grow, and MRLS can only shrink.
Thus, the number of iterations is bounded by the size of $\mathcal{L}$.
The iteration terminates almost always in practice. In most programs,
$\mathcal{L}$ is finite, and the termination is guaranteed. However, some
programs have a recursive data structure with locks, which makes $\mathcal{L}$
infinite. That said, a recursive function interacting with an unbounded number
of locks is rare in practice, so the iteration can terminate despite
$\mathcal{L}$ being infinite.

We can easily generalize this approach to mutually recursive functions. Given a
set of mutually recursive functions, $f_1,\cdots,f_n$, we set all the MELSs and
MRLSs to the bottom values: $\dvar{MELS}{}{f_i}=\emptyset$ and
$\dvar{MRLS}{}{f_i}=\mathcal{L}$. We then analyze each function and update them
with the results. We repeat the analysis until none of them change.

\vspace{-0.25em}
\subsection{Top-Down Data Fact Propagation}
\label{sec:top-down}

A function summary for the transformation has to contain the \emph{entry lock
set} (ELS) and the \emph{return lock set} (RLS) of the function. They are locks
that can be always held at the entry and the return, respectively. It is important that they
are different from the MELS and MRLS. The ELS contains guards given to a function by
its caller, and the MELS contains some of them, which are dropped by the function.
Consequently, the MELS is always a subset of the ELS. Similarly, the RLS contains
guards returned by a function to its caller, and the MRLS contains some of them, which
are constructed in the function. The MRLS is always a subset of the RLS. In
addition, the following equation holds:
$\dvar{ELS}{}{}-\dvar{MELS}{}{}=\dvar{RLS}{}{}-\dvar{MRLS}{}{}$. We call this
common difference the \emph{propagated lock set} (PLS). The PLS of a function is
the set of guards given from and returned to its caller.

\begin{example}
Both MELS and MRLS of \C{inc} are $\emptyset$, but both ELS and RLS of \C{inc}
are $\{\C{m}\}$.
\vspace{-0.25em}
\begin{lstlisting}[language=Rust]
fn safe_inc() {
    pthread_mutex_lock(&mut m); inc();
    pthread_mutex_unlock(&mut m);
}
fn inc() { n += 1; }
\end{lstlisting}
\vspace{-0.5em}
We need the ELS and RLS of \C{inc} to identify the data-lock relation correctly.
If we consider only the MELS and MRLS, we incorrectly conclude that \C{m}
does not protect \C{n}.
\end{example}

The goal of the top-down data fact propagation is to compute the ELS and RLS of each
function. We first compute the ELS of each function. It allows us to find the
PLS by subtracting the MELS from the ELS. Then, the union of the PLS and the
MRLS is the RLS.

We first collect all the available guards at each function call. The
arguments of the function call are collected together to recompute paths
according to aliasing. $\dvar{Call}{}{f,g}$ is the set of pairs, each of which
consists of the set of available guards and the list of arguments when $f$ calls
$g$. Because $f$ may call~$g$ multiple times, multiple pairs may exist.
Available guards at each call are already computed during AGA.
Thus, $\dvar{Call}{}{f,g}$ is:
\vspace{-0.75em}
\[
\small
  \dvar{Call}{}{f,g}=\{(\dvar{In}{A}{s},\rep{e})\ |\ s=g(\rep{e})\ \land\ s\
  \text{is in}\ f\}
\vspace{-0.5em}
\]

We then perform a top-down dataflow analysis to compute the ELS of each
function. The analysis is cheap because it does not
analyze function bodies and simply propagates data facts through call edges.  If
a function does not have any callers, its ELS is the same as its MELS.
Otherwise, its callers propagate available guards. Each caller propagates not
only the available guards identified by AGA, but also the guards propagated from
its own callers. We want always-propagated guards, so we compute the
intersection of the guards from each caller. The dataflow equations are as
follows:
\vspace{-0.5em}
\[
\small
\begin{array}{r@{~}c@{~}l}
  \dvar{ELS}{}{g} &=& \begin{cases}
    \dvar{MELS}{}{g} & \text{if}\ \dvar{Pred}{}{g} = \emptyset \\
    \bigcap_{f\in\dvar{\scriptsize Pred}{}{g}}\dvar{Prop}{}{f,g} & \text{otherwise}
  \end{cases} \\
  \dvar{Prop}{}{f,g} &=&
  \bigcap_{(P,\rep{e})\in\dvar{\scriptsize Call}{}{f,g}}\alias(P\cup\dvar{ELS}{}{f},\rep{e},\dvar{Params}{}{g})
\end{array}
\vspace{-0.5em}
\]
where $\dvar{Prop}{}{f,g}$ is the set of guards propagated from $f$ to $g$.

We finally compute the PLS and RLS:
\vspace{-0.5em}
\[
\small
\begin{array}{r@{~}c@{~}lr@{~}c@{~}l}
  \dvar{PLS}{}{} &=& \dvar{ELS}{}{}-\dvar{MELS}{}{} & \qquad
  \dvar{RLS}{}{} &=& \dvar{MRLS}{}{}\cup\dvar{PLS}{}{} \\
\end{array}
\vspace{-0.5em}
\]

\addtocounter{example}{-1}

\begin{example} (\emph{continued})
  \begin{itemize}
    \item $\dvar{Call}{}{\texttt{safe\_inc},\texttt{inc}}=\{(\{\C{m}\},[\,])\}$
    \item $\dvar{Prop}{}{\texttt{safe\_inc},\texttt{inc}}=\{\C{m}\}$
    \item $\dvar{ELS}{}{\texttt{inc}}=\dvar{PLS}{}{\texttt{inc}}=\dvar{RLS}{}{\texttt{inc}}=\{\C{m}\}$
  \end{itemize}
\end{example}

After finishing the top-down propagation, we can generate a function summary of each
function. \el and \rl are the same as the ELS and RLS, respectively. The set of
locks held in each line, required by \lline, is determined by combining
available guards identified by AGA and the PLS of the function.

\subsection{Data-Lock Relation Identification}
\label{sec:data-lock}

We identify the data-lock relation from the flow-lock relation computed by the
preceding analysis. The key idea is to find the lock held at each access to a
certain path. Since the global pattern (\cref{sec:global}) and the struct
pattern (\cref{sec:struct}) require different treatments, we split
paths into global variables and struct fields and compute the data-lock relation
of each.

We first discuss the global pattern. The first step is to collect every access
to each global variable. We record all the available guards and whether the
access is read or write. The available guards are the union of those found by
AGA and the PLS of the function where the access happens. $\dvar{Acc}{}{x}$ is
the set of accesses to a global variable $x$:
\[
\small
\dvar{Acc}{}{x}=\{(s,\dvar{In}{A}{s}\cup\dvar{PLS}{}{f},a)\ |\ \embox{access}(s,x,a)
\ \land\ s\ \text{is in}\ f\}
\]
where $\embox{access}(s,x,\dvar{r}{}{})$ and
$\embox{access}(s,x,\dvar{w}{}{})$ hold
when $s$ reads $x$ and $s$ modifies $x$, respectively.
We then find a candidate lock for each global variable. A candidate lock is a
lock that is held most frequently when accessing the variable:
\[
\small
\dvar{Cand}{}{x}=\argmax_y |\{(s,P,a) \in \dvar{Acc}{}{x}\ |\ y\in P\}|
\]
We split accesses into safe and unsafe ones according to the existence of the
candidate lock. We consider an access safe if it happens when the candidate
is held, and unsafe otherwise:
\[
\small
\begin{array}{r@{~}c@{~}l}
  \dvar{Safe}{}{x} &=& \{(s,P,a) \in \dvar{Acc}{}{x}\ |\ \dvar{Cand}{}{x}\in P\}
  \\
  \dvar{Unsafe}{}{x} &=& \{(s,P,a) \in \dvar{Acc}{}{x}\ |\ \dvar{Cand}{}{x}\not\in P\}
\end{array}
\]

To determine the data-lock relation, we need to check whether each statement is
\emph{concurrent}, \ie, can run concurrently with other threads.  The existence
of an unsafe access does not necessarily mean that the candidate lock does not
protect the global variable.  If a statement is non-concurrent, it can safely
access a global variable without holding a lock.  Therefore, the precise
identification of the data-lock relation requires a precise thread analysis.
Since a precise thread analysis is expensive, we instead propose a simple
heuristic. We consider a statement concurrent only if it belongs to a
function reachable from an argument to \C{pthread\_create}, the thread-spawning
function.

Using the heuristic, we determine whether a candidate lock really protects the
global variable. The candidate protects the variable if a safe write access
exists and every unsafe access happens in a non-concurrent statement.
\[
\small
\begin{array}{l}
\dvar{Cand}{}{x}\ \text{protects}\ x\ \text{iff}\ \\
(\exists(s,P,a)\in\dvar{Safe}{}{x},a=\dvar{w}{}{})
\ \land \\
(\forall(s,P,a)\in\dvar{Unsafe}{}{x},s\ \text{is non-concurrent})
\end{array}
\]

For the struct pattern, we collect accesses to each field of a struct. A
candidate lock for a field must be a field in the same struct.
$\dvar{Acc}{}{T,l}$ is the set of accesses to a field $l$ in a type $T$, and
$\dvar{Cand}{}{T,l}$ is a candidate for it:
\[
\small
\begin{array}{r@{~}c@{~}l}
  \dvar{Acc}{}{T,l} &=& \{(s,\dvar{In}{A}{s}\cup\dvar{PLS}{}{f},p,a)\ | \\
  &&\embox{access}(s,p.l,a)
\ \land\ \embox{type}(p)=T\ \land\ s\ \text{is in}\ f\} \\
  \dvar{Cand}{}{T,l} &=& \argmax_{l'} |\{(s,P,p,a) \in \dvar{Acc}{}{x}\ |\ p.l'\in P\}|
\end{array}
\]

We split accesses into safe and unsafe ones and check whether the candidate
protects the field, just as we do for the global pattern. The only difference is
that the condition for a statement to be considered concurrent is stricter than
before. A struct value is not accessible from other
threads right after its creation. It becomes accessible only after the function shares it with other
threads by storing it in a global data structure or passing it as a thread
argument. Determining when a value is shared requires a precise
thread analysis as well, so we propose a heuristic. For a given struct value containing
a lock field $l$, we consider a statement non-concurrent not only when its
function is unreachable from an argument to \C{pthread\_create}, but also when
the function initializes $l$ by calling \C{pthread\_mutex\_init}. Such a function
is usually where the struct value is created and uniquely accessed.

%% file: evaluation.tex
\section{Evaluation}
\label{sec:evaluation}

\newcommand{\SUCC}{\ding{51}}
\newcommand{\FAIL}{\ding{55}}
\newcommand{\SUCF}{\FAIL\ding{222}\SUCC}

\begin{figure}
  \centering
  \includegraphics[width=0.45\textwidth]{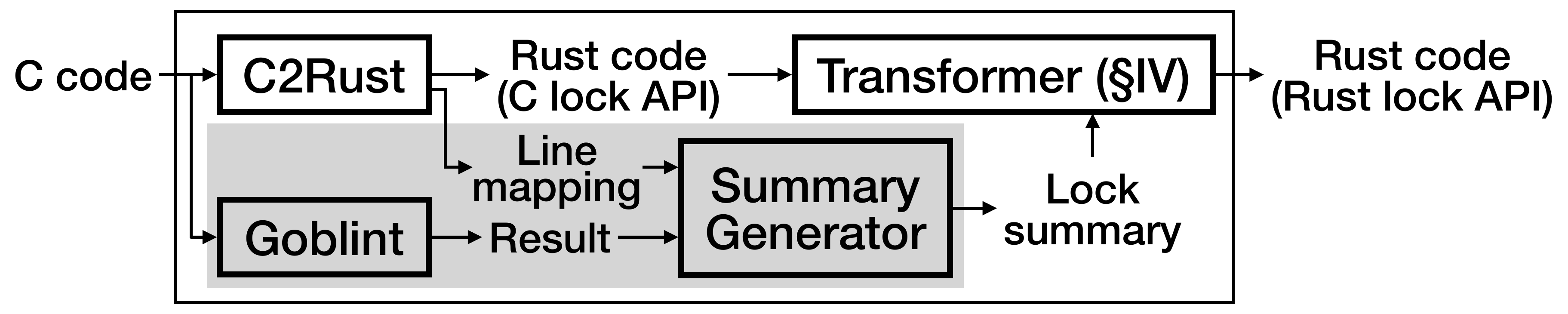}
  \caption{Workflow of Concrat$_G$ (gray: differences from Concrat)}
  \label{fig:goblint}
\vspace*{-1em}
\end{figure}

\begin{table*}[t]
\caption{Test set and experimental results}\label{tab:evaluation}
\vspace*{-.5em}
\centering
\scriptsize
\renewcommand{\arraystretch}{.95}
\begin{tabular}{
    @{~}c@{~}||
    @{~}r@{~}|@{~}r@{~}|@{~}r@{~}|@{~}r@{~}|@{~}r@{~}|@{~}r@{~}||
    @{~}r@{~}|@{~}r@{~}|@{~}r@{~}|@{~}c@{~}|@{~}c@{~}|@{~}c@{~}|@{~}c@{~}|@{~}c@{~}|@{~}c@{~}||
    @{~}r@{~}|@{~}r@{~}|@{~}c@{~}
}
  & \multicolumn{6}{@{~}c@{~}||@{~}}{Size} &
  \multicolumn{9}{@{~}c@{~}||@{~}}{Transformation} &
  \multicolumn{3}{@{~}c@{~}}{Analysis} \\ \hline
  Project &
  C LOC & Rust LOC & Mutex & Rwlock & Spin & Cond &
  Time & Insertion & Deletion & Succ & Reason & Fix & Test & Test$_\text{C}$ & Test$_\text{O}$ &
  Time$_\text{O}$ & \multicolumn{1}{@{~}c@{~}|@{~}}{Time$_\text{G}$} & Succ$_\text{G}$ \\ \hline
AirConnect&17516&32565&88&0&0&14&1.159&870&874&\FAIL&cond acq&&&&&1.701&timeout&\\
axel&5848&7685&16&0&0&0&0.344&78&112&\SUCC&&&&&&0.409&error&\\
brubeck&5635&6769&15&0&17&0&0.289&99&91&\SUCC&&&\SUCC&\SUCC&\SUCC&0.362&80.393&\FAIL\\
C-Thread-Pool&710&791&23&0&0&7&0.059&106&113&\SUCC&&&\SUCC&\SUCC&\SUCC&0.060&0.728&\SUCC\\
cava&4768&6538&10&0&0&0&0.289&29&27&\SUCC&&&&&&0.334&72.908&\SUCC\\
Cello&20885&30796&5&0&0&0&1.521&27&15&\FAIL&func ptr&&&&&3.103&error&\\
Chipmunk2D&16053&21509&12&0&0&10&0.839&52&60&\SUCC&&&&&&2.074&error&\\
clib&25073&66287&38&0&0&0&2.416&193&207&\FAIL&func ptr&&&&&4.234&timeout&\\
dnspod-sr&9259&12596&0&0&99&0&0.537&272&234&\SUCF&&dead&&&&0.696&1667.203&\FAIL\\
dump1090&4646&6281&9&0&0&6&0.259&44&77&\SUCC&&&\SUCC&\SUCC&\SUCC&0.307&timeout&\\
EasyLogger&2011&29298&4&0&0&0&0.219&428&415&\FAIL&cond acq&&&&&0.234&6.458&\FAIL\\
fzy&2621&4013&4&0&0&0&0.154&16&16&\SUCC&&&\SUCC&\SUCC&\SUCC&0.168&error&\\
klib&716&1016&14&0&0&14&0.076&62&100&\SUCC&&&\SUCC&\SUCC&\SUCC&0.078&error&\\
kona&38850&48583&10&0&0&0&2.067&37&31&\SUCC&&&\SUCC&\SUCC&\SUCC&3.223&timeout&\\
level-ip&5414&6651&36&23&0&4&0.329&232&317&\FAIL&func ptr&&&&&0.442&timeout&\\
libaco&1282&1800&6&0&0&0&0.090&22&33&\SUCC&&&\SUCC&\SUCC&\SUCC&0.105&timeout&\\
libfaketime&521&806&6&0&0&6&0.059&43&85&\SUCC&&&\SUCC&\SUCC&\FAIL&0.059&0.147&\SUCC\\
libfreenect&627&962&10&0&0&4&0.066&87&116&\SUCC&&&&&&0.069&3.439&\SUCC\\
libqrencode&6670&9013&4&0&0&0&0.367&28&40&\SUCC&&&\SUCC&\SUCC&\SUCC&0.447&error&\\
lmdb&10827&16290&27&0&0&6&0.722&266&292&\FAIL&lock arg&&&&&0.910&error&\\
minimap2&17279&23531&6&0&0&4&1.044&26&56&\SUCC&&&\SUCC&\SUCC&\SUCC&1.438&73.902&\SUCC\\
Mirai-Source-Code&1839&2889&14&0&0&0&0.151&118&135&\SUCF&&goto&&&&0.164&43.736&\SUCF\\
neural-redis&3645&6312&12&0&0&0&0.261&51&59&\SUCC&&&&&&0.310&186.593&\SUCC\\
nnn&12091&16424&7&0&0&0&0.822&37&66&\SUCC&&&&&&1.056&2264.742&\SUCC\\
pg\_repack&7420&8152&10&0&0&0&0.306&36&35&\SUCC&&&\SUCC&\SUCC&\SUCC&0.384&error&\\
phpspy&19390&29860&8&0&0&10&1.199&1441&1479&\FAIL&cond acq&&&&&1.580&error&\\
pianobar&11452&33212&45&0&0&17&1.248&132&182&\SUCC&&&&&&1.624&timeout&\\
pigz&9118&12660&5&0&0&7&0.471&176&178&\SUCF&&no ret&\SUCC&\SUCC&\SUCC&0.615&error&\\
pingfs&2318&3332&26&0&0&6&0.180&110&137&\FAIL&cond acq&&&&&0.198&9.278&\FAIL\\
ProcDump-for-Linux&4152&6961&31&0&0&11&0.245&171&438&\SUCC&&&\FAIL&&&0.286&79.284&\FAIL\\
proxychains&2686&5460&6&0&0&0&0.218&44&52&\SUCC&&&&&&0.223&49.203&\SUCC\\
proxychains-ng&5203&9031&8&0&0&0&0.389&24&32&\SUCC&&&&&&0.444&1058.823&\SUCC\\
Remotery&7212&9361&4&0&0&0&0.397&34&24&\FAIL&cond acq&&&&&0.567&error&\\
sc&142&206&7&0&0&0&0.029&34&54&\SUCC&&&\SUCC&\SUCC&\SUCC&0.030&0.049&\SUCC\\
shairport&8605&12533&37&0&0&0&0.576&1093&1046&\FAIL&cond acq&&&&&0.736&error&\\
siege&19281&25412&21&0&0&16&1.053&202&293&\FAIL&lock arg&&&&&1.882&error&\\
snoopy&2262&4605&4&0&0&0&0.198&23&38&\SUCC&&&\SUCC&\SUCC&\SUCC&0.234&3.855&\SUCC\\
sshfs&7193&9914&75&0&0&13&0.388&277&302&\SUCC&&&\SUCC&\SUCC&\SUCC&0.517&2028.390&\FAIL\\
streem&20169&31444&36&0&0&0&1.070&115&114&\SUCF&&bug fix&\SUCC&\SUCC&\SUCC&2.084&error&\\
stud&7931&10789&12&0&0&0&0.423&61&54&\SUCC&&&&&&0.505&error&\\
sysbench&16020&41222&19&10&0&9&0.999&110&244&\SUCF&&goto&\SUCC&\FAIL&&1.149&error&\\
the\_silver\_searcher&7242&12453&23&0&0&5&0.441&112&177&\SUCC&&&\SUCC&\SUCC&\SUCC&0.548&error&\\
uthash&817&1450&0&6&0&0&0.078&56&57&\SUCC&&&\SUCC&\SUCC&\SUCC&0.091&0.100&\SUCC\\
vanitygen&10919&9710&23&0&0&11&0.379&121&224&\SUCC&&&&&&0.454&error&\\
wrk&8658&12255&12&0&0&0&0.465&35&42&\SUCC&&&&&&0.525&error&\\
zmap&17435&24366&12&0&0&0&0.795&42&71&\FAIL&lock arg&&&&&1.257&timeout&\\
\end{tabular}
\\[0.5em]
\scriptsize{(Times are in seconds.
Subscript C means C2Rust, O means ours, and G means Goblint.)}
\vspace*{-1em}
\end{table*}

Our experiments are on an Ubuntu machine with Intel Core
i7-6700K (4 cores, 8 threads, 4GHz) and 32GB DRAM.

\subsection{Implementation}
\label{sec:implementation}

We implemented Concrat on top of the Rust
compiler~\cite{rustc}. The transformer lowers given code to the compiler's high-level
intermediate representation~\cite{hir} and walks it to replace the C lock API.
The analyzer uses the compiler's
dataflow analysis framework~\cite{rustc-dataflow} for its mid-level intermediate
representation~\cite{mir}.

Concrat handles not only mutexes, but also read-write locks, spin locks, and
condition variables.
Since Rust recommends using mutexes instead of spin locks~\cite{rust-spinlock},
we replace them with
\C{RwLock}~\cite{rwlock}, \C{Mutex}, and \C{Condvar}~\cite{condvar} of \C{std::sync}.
Concrat does not support re-entrant locks because the Rust standard library does
not provide them.

To compare our analyzer with the state-of-the-art static analyzer for concurrent
programs, we additionally built Concrat$_G$. \cref{fig:goblint} illustrates the
workflow of Concrat$_G$. It uses Goblint~\cite{goblint,goblint21} to analyze C
code. Goblint computes both data-lock and flow-lock relations using abstract
interpretation~\cite{ai}. Since Goblint's result contains line numbers for C
code, our summary generator replaces them with those for Rust code using the
C-to-Rust line mappings generated by C2Rust.

Note that we can change the implementation of Concrat to analyze C code,
instead of Rust code, just like Concrat$_G$, because the proposed analysis is
language-agnostic. Our choice of analyzing Rust code eases implementation as we
can utilize the Rust compiler's dataflow analysis framework.

\subsection{Test Set Collection}

We collected 46 real-world concurrent C programs,
all of the public GitHub repositories satisfying the following conditions:
1) more than 1,000 stars, 2) not a study material, 3) using the pthread lock API at least once,
4) C code less than 500,000 bytes, and 5) translatable with C2Rust.
Two projects satisfied the first four conditions but not the last; they
use C11 Atomics, but C2Rust supports only C99-compliant code.
When C2Rust generates uncompilable code due to wrong type casts,
we included such projects after manually fixing them.

The first seven columns of \cref{tab:evaluation} show the collected programs
and the code size of each. The second and third columns show the lines of C code
and C2Rust-generated code, respectively; the fourth to seventh columns show the
numbers of \C{pthread\_mutex\_*}, \C{pthread\_rwlock\_*}, \C{pthread\_spin\_*},
and \C{pthread\_cond\_*} calls, respectively.

\subsection{Transformation}

We evaluate our transformer with the following questions:
\begin{itemize}
  \item RQ1. Scalability: Does it transform large programs in a reasonable amount of time?
  \item RQ2. Applicability: Does it handle most code patterns found in real-world programs?
  \item RQ3. Correctness: Does it preserve the semantics of the original program?
\end{itemize}

\subsubsection{Scalability}

We translate the programs with Concrat to evaluate the transformer's scalability.
In \cref{tab:evaluation},
the eighth column shows the transformation time; the ninth and tenth
show the inserted and deleted lines, measured with \C{diff}.

\begin{figure}
  \centering
  \includegraphics[width=0.4\textwidth]{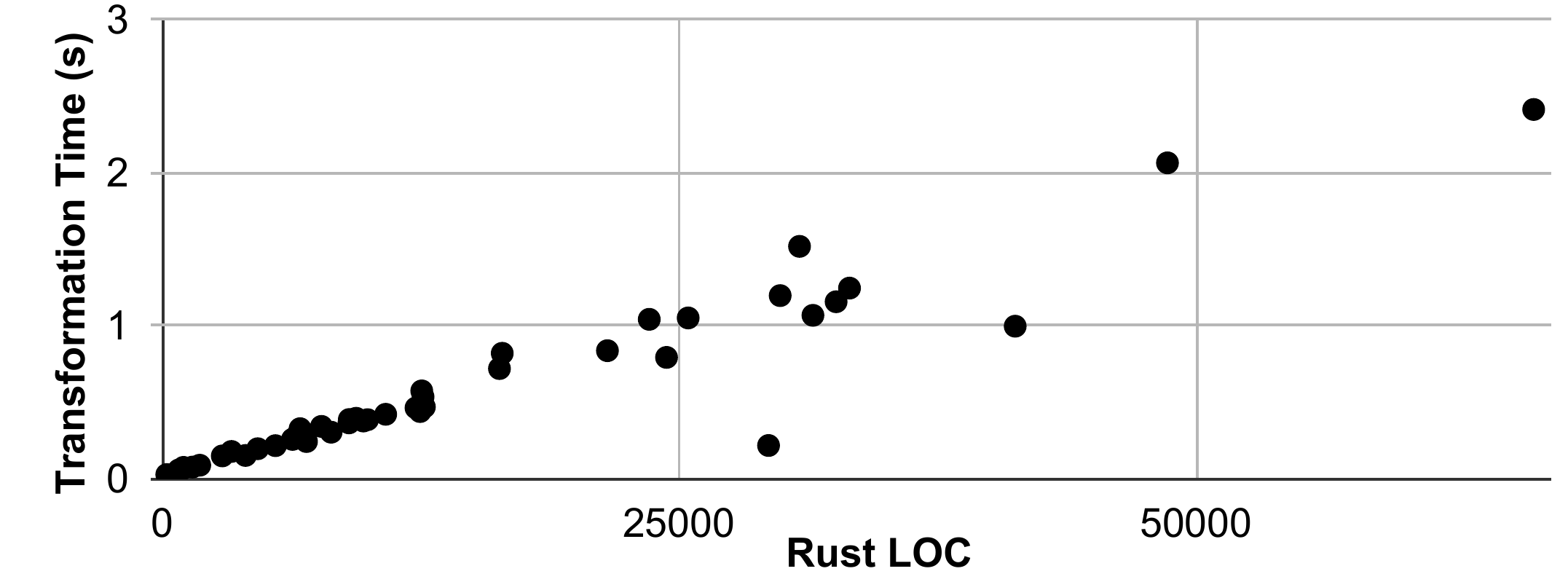}
  \caption{Transformation time according to Rust LOC}
  \label{fig:transformation}
\vspace*{-1em}
\end{figure}

The result shows that the transformer is scalable. As \cref{fig:transformation}
shows, the transformation time is proportional to the Rust LOC, and it takes
less than 2.5 seconds to transform 66 KLOC by inserting and deleting hundreds of
lines.

\subsubsection{Applicability}

We check whether the transformer handles diverse code patterns in
real-world programs to evaluate its applicability. We consider that the
transformer successfully handles a certain pattern if the transformed code is
compilable.  In \cref{tab:evaluation},
the eleventh column shows compilability; the twelfth shows
the reason for a failure; the thirteenth shows our manual fix for the
original code to make compilation succeed.

The transformer has high applicability. Among 46,
29 are compilable, 5 are compilable requiring manual fixes, and
12 are not.  Overall, 74\% of the programs are compilable.

\paragraph{Failures}

We manually investigated the reasons for compilation failures and found three code patterns.
\begin{itemize}[leftmargin=*]
  \item \emph{Conditional acquisitions (cond acq)}:
A function conditionally acquires a lock. Consider the following code:
\begin{lstlisting}[language=Rust]
fn may_lock() -> i32 {
    if b {pthread_mutex_lock(&mut m); 0} else {1}
}
if may_lock() == 0 {
    n += 1; pthread_mutex_unlock(&mut m);
}
\end{lstlisting}
The function \C{may\_lock} acquires a lock and returns \C{0} if a certain
condition is satisfied, and otherwise returns \C{1} without acquiring the lock.
Its caller accesses the shared data only when the return value is \C{0}.
The transformer cannot handle this pattern. Since \C{may\_lock} has empty \rl,
it does not return any guards after the transformation. Its caller drops an
unowned guard, thereby being uncompilable. To solve this, we have to make
\C{may\_lock} return \C{Option}~\cite{option} of a guard.
Conditionally-succeeding functions appear in most C programs, not only
in concurrent ones. Translating them to functions returning \C{Option} will
be promising future work.

%

\item \emph{Function pointers (func ptr)}:
A function that takes or returns guards is used as a function pointer. Since
adding guards changes the type of the function pointer, the transformed code
is uncompilable.
To address this pattern, we need to transform functions that take
function pointers. Such functions are often in a library; if it is the case,
the library also should be rewritten in Rust.

\item \emph{Lock arguments (lock arg)}:
A function takes only a pointer to a lock as an argument.
It would be an interesting improvement to handle this pattern with
generic functions. If the function does not access any data protected by a
specific lock, we can make the function take an argument of \C{Mutex<T>},
where \C{T} is a type parameter.

\end{itemize}

\paragraph{Manual Fixes}

We made four kinds of manual fix.
We first explain two code patterns requiring manual fixes.
\begin{itemize}[leftmargin=*]
\item \emph{Removing \C{goto} (goto)}:
A function uses \C{goto}.
\begin{lstlisting}[language=C]
if (b) { pthread_mutex_unlock(&m); goto err; }
n += 1; ... return 0;
err: return 1;
\end{lstlisting}
For the above code, C2Rust replaces \C{goto} with \C{if}:
\begin{lstlisting}[language=Rust]
if b { pthread_mutex_unlock(&mut m); x = 123; }
if x != 123 { n += 1; ... return 0; }
return 1;
\end{lstlisting}
Then, the transformer generates uncompilable code:
\begin{lstlisting}[language=Rust]
if b { drop(m_guard); x = 123; }
if x != 123 { *m_guard += 1; ... return 0; }
return 1;
\end{lstlisting}
Since type checking is path-insensitive, it considers \C{m\_guard} possibly
unowned in the second line.
We replaced \C{goto} with the statements after the jump target:
\begin{lstlisting}[language=C]
if (b) { pthread_mutex_unlock(&m); return 1; }
n += 1; ... return 0;
\end{lstlisting}
which Concrat translates to compilable code.
\item \emph{Changing return type of no-return function (no ret)}:
  A function does not return. Consider the following code:
\begin{lstlisting}[language=Rust]
fn err() { exit(-1); }
if b { pthread_mutex_unlock(&mut m); err(); }
n += 1; ...
\end{lstlisting}
Since \C{exit} does not return, \C{err} does not either and
accessing \C{n} is safe. But, the type checker does not
know that \C{err} never returns, and the transformed code is uncompilable:
\begin{lstlisting}[language=Rust]
if b { drop(m_guard); err(); }
*m_guard += 1; ...
\end{lstlisting}
The type checker considers \C{m\_guard} possibly unowned in the last line.
We changed the return type of \C{err} to \C{!}~\cite{never-type}:
\begin{lstlisting}[language=Rust]
fn err() -> ! { ... exit(-1); }
\end{lstlisting}
indicating no return. The transformed code is compilable.
\end{itemize}
Both manual fixes can be avoided by improving C2Rust.
We now explain program-specific fixes.
\begin{itemize}[leftmargin=*]
\item \emph{Removing dead code (dead)}:
  We deleted a function taking a pointer to a lock because it is never called.
\item \emph{Bug fix (bug fix)}:
streem has the following code (simplified):
\begin{lstlisting}[language=C]
pthread_mutex_unlock(&m);
if (...) { pthread_mutex_unlock(&m); return; }
\end{lstlisting}
Due to the second \C{pthread\_mutex\_unlock} call, the transformed code is
uncompilable. We believe it is a bug and removed it.
We contacted the developer but have not received a response yet.
This confirms the common belief that rewriting legacy programs in Rust can
reveal unknown bugs.
\end{itemize}

\subsubsection{Correctness}

We ran the test cases of each program whose transformation succeeds to
evaluate the correctness of the transformer. A correct transformer must preserve
the semantics of the original program. We consider the transformer correct if
the transformed program passes all of its test cases. The fourteenth to
sixteenth columns show whether the original C program, the C2Rust-generated
program, and the transformed program pass the test cases, respectively.

The result shows that our approach transforms most programs correctly. Among 34
programs compilable after the transformation possibly with manual fixes, 14 have
no test cases or only those covering no lock API calls, so we performed the
evaluation with the remaining 20. One fails even before C2Rust's translation. One
fails after C2Rust's translation because it incorrectly translates some inline
assembly. After the transformation, 17 still pass their tests, but 1 fails.

The failing program does not reveal an inherent limitation of our approach. The
reason for the incorrect transformation is the imprecise \C{timespec} tracking
of our transformer implementation. While \C{pthread\_cond\_timedwait} of the C
lock API takes what time to wait for, its Rust counterpart takes how long to
wait. To address this discrepancy, the transformer syntactically finds a
\C{clock\_gettime} call, which sets a given \C{timespec} to the current time,
and how many seconds are added to the \C{timespec} before calling
\C{pthread\_cond\_timedwait}. However, the failing program uses multiple
\C{timespec} values, whose relation cannot be found by our syntactic analysis.
We can easily resolve this issue by adopting intraprocedural value analysis for
\C{timespec}.

It is not surprising that the transformer is correct in most cases as far as the
transformed code is compilable because the design of the transformation
justifies the correctness. It transforms a \C{lock} function call to a \C{lock}
method call and an \C{unlock} function call to the drop of a guard whose
finalizer unlocks the connected lock. Since the names of lock and guard
variables are syntactically related, the dropped guard always unlocks the
correct lock. It transforms each lock-protected data access to field access
through a guard. Again, the lock and guard names are syntactically related, so
the transformed code always accesses the correct data. The only possible threat
is guards being used before initialization or after destruction, but the
compiler detects it.



\subsection{Analysis}

We evaluate our analyzer with the following questions:
\begin{itemize}
  \item RQ4. Scalability: Does it analyze large programs quickly, compared to the
    state-of-the-art static analyzer?
  \item RQ5. Precision: Does it produce precise lock summaries, compared to the
    state-of-the-art static analyzer?
\end{itemize}

\subsubsection{Scalability}

We translate the collected programs with both Concrat and Concrat$_G$ and
measure the analysis time to compare the scalability of our analyzer and
Goblint. For Goblint, we use the
\C{medium-program.json} configuration~\cite{goblint-conf} and
additionally enable the \C{allfuns} option to analyze
every function for programs without \C{main}.
It makes Goblint perform flow-, context-, path-sensitive analysis.
We set a 24-hour time limit. The seventeenth and eighteenth
columns of \cref{tab:evaluation} show the time required by ours and Goblint.

The result shows that our analyzer is more scalable than Goblint. Goblint fails
to analyze 19 programs due to internal errors.
Our analyzer processes all the remaining 27 programs faster
than Goblint. It takes less than 4.3 seconds to analyze 66 KLOC. On the other
hand, Goblint does not even finish the analysis of eight programs in the time limit
and takes 1.1$\times$ to 3923$\times$ more than ours to analyze the other 19 programs.

\subsubsection{Precision}

We measure the precision of lock summaries generated by our analyzer and Goblint
to compare their precision. We use the compilability of code translated by Concrat
and Concrat$_G$ as a proxy for assessing the analyzers' precision because
an imprecise flow-lock relation makes transformed code uncompilable, as discussed
in \cref{sec:analysis}.
The eleventh and nineteenth columns of \cref{tab:evaluation} show compilability
after Concrat's and Concrat$_G$'s translation, respectively.

The result shows that our analyzer is more precise than Goblint for
four programs, generating summaries that made the transformed code compilable.
Our manual investigation confirms that those translated by Concrat$_G$
are uncompilable due to imprecision in the flow-lock relations.
Goblint's imprecision mainly stems from the imprecise lock-aliasing
information. It knows locks \C{a} and \C{b} are aliased when \C{a} is locked,
but this information is sometimes lost due to overapproximation, and \C{b} is
considered still locked even after \C{a} is unlocked, leading to imprecise
flow-lock relations.

\subsection{Threats to Validity}

The primary threats to internal validity lie in the implementation of
our tool. We implemented it with the dataflow analysis framework of the Rust
compiler, which is already massively used and tested by the compiler.

The threats to external validity concern the choice of the translated C
projects, each of which has more than 1,000 stars and C code of fewer than
500,000 bytes. Less popular or larger projects may have code patterns unseen in
the selected projects. Further experiments with more C projects can give more
confidence in the generalizability of our approach.

The threats to construct validity include evaluation metrics. We used whether a
compilation succeeds or not as a proxy for the applicability of the transformer
and the precision of the analyzer. We ran test cases to evaluate the correctness
of the transformer. While test cases cannot guarantee the semantics preservation
of the transformation, in practice, running test cases is the most popular way
to check whether a certain program has the intended semantics.

%% file: related.tex
\section{Related Work}
\label{sec:related}

\subsubsection{Transforming C2Rust-Generated Code}

A few tools have been proposed to replace C features in C2Rust-Generated code
with their counterparts in Rust, but they do not target the lock API.
Emre et al.~\cite{safer-rust} tackled replacing C pointers
in Rust with Rust pointers. Their work also requires semantic information,
but they focus on pointers and sequential programs.
Ling et al.~\cite{crusts} proposed CRustS, which contains 220 syntactic
replacement rules written in the TXL transformation language~\cite{txl}.
It targets simpler features than this work.

\subsubsection{C to Safe Languages}

Many safe substitutes for C and (semi-) automatic translation to those languages
have been
proposed~\cite{ccured,cyclone,3c,checked-c,deputy,cqual,ccc,safe-c,c-at,rc-compiler}.
However, they guarantee only some form of
memory safety and do not provide any safe lock API.
Necula et al.~\cite{ccured} proposed CCured, a language extending C with new
kinds of pointer and the type system to statically verify or dynamically
enforce the safety of each pointer.
%
Grossman et al.~\cite{cyclone} proposed Cyclone, which extends C with
region-based memory management~\cite{region-based}. Cyclone's type system
prevents dangling pointer dereference by annotating each pointer with a region
to use the pointer.
%
Machiry et al.'s {\bf 3C} tool~\cite{3c} automatically translates C to Checked
C~\cite{checked-c}, an extension of C with checked pointer types.
 Checked pointers guarantee the absence
of null pointer dereference and out-of-bound accesses.

\subsubsection{Static Analysis for Concurrent Programs}
Various static analyzers for concurrent C programs have been proposed. Our
analyzer analyzes concurrent Rust programs but assumes only C2Rust-generated code.
%
While the goal of existing analyzers is to detect concurrency bugs,
our goal is to efficiently generate lock summaries
for our transformation. This different goal makes our design
completely distinct from the others. Our analyzer targets the same precision as
the Rust type checker by using context- and path-insensitive dataflow analyses.
However, other analyzers perform context- or path-sensitive analyses to
reduce false alarms. Goblint~\cite{goblint,goblint21} and
others~\cite{astree,mine11,mine14} are based on abstract
interpretation~\cite{ai}.
RELAY~\cite{relay} and SDRacer~\cite{sdracer} utilize symbolic
execution~\cite{symbolic-execution}.
Locksmith~\cite{locksmith} and Kahlon et al's tool~\cite{cobe} perform
context-sensitive analyses.

A few static analyzers for concurrent Java programs~\cite{sword,racerd} exist.
Because Java provides a syntactic \C{synchronized} block,
a lock acquired by a certain
function cannot be released by another function, which frequently happens in C.

\subsubsection{API Mapping Mining}

Researchers have proposed API mapping
mining~\cite{10.1145/1806799.1806831,6227179,10.1145/2642937.2643010}, which
automatically discovers syntactic discrepancies, e.g., type/function/method
names and receiver/parameter positions, between the APIs of different languages.
Such information can be utilized by syntactic translators. It is useful when
there are many API functions, and their mappings are unknown.

Unfortunately, API mapping mining is ineffective in dealing with our problem. We
focus on translating concurrent programs, which significantly rely on the lock
API. The lock API provides only a few functions, and their mappings can be
constructed with low manual effort. Our challenge rather resides in the semantic
discrepancies between C and Rust’s lock APIs. The Rust lock API requires
information not written in C code to be explicitly written. This work proposes a
dataflow analysis technique to find the required information.

\subsubsection{Learning from Code Edit Examples}

To facilitate automatic code transformation, researchers have proposed
techniques to learn code edit rules from code edit
examples~\cite{6606596,8813263}. However, these example-based approaches
do not fit our problem, which requires code transformation that cannot be syntactically generalized.
For example, to transform a lock-initializing API call, we need the name of every variable
protected by the lock. While code edit examples contain variable names, the
names originate from arbitrary source code locations, and we cannot
syntactically generalize the examples to code edit rules regardless of the number of examples.

%% file: conclusion.tex
\section{Conclusion}
\label{sec:conclusion}

We tackle the automatic C-to-Rust translation of concurrent programs
by replacing the C lock API with the Rust lock API utilizing
data-lock and flow-lock relations. Our solution is a transformer
replacing the lock API using a lock summary efficiently and precisely generated
by our static analyzer. Our evaluation shows that
the transformer is scalable, widely applicable, and correct, and the analyzer is
scalable and precise.